# Competition of chiral soliton lattice and Abrikosov vortex lattice in QCD with isospin chemical potential

**Martin Spillum Grønli**[1,a] **, Tomáš Brauner**[2,b]

[1] Department of Physics, Norwegian University of Science and Technology, 7491 Trondheim, Norway
[2] Department of Mathematics and Physics, University of Stavanger, 4036 Stavanger, Norway

**Abstract** We investigate the thermodynamics of two-flavor quark matter in presence of nonzero isospin chemical potential and external magnetic field. It is known that at sufficiently high isospin chemical potential, charged pions undergo Bose–Einstein condensation (BEC). The condensate behaves as a superconductor, exhibiting Meissner effect in weak external magnetic fields. Stronger fields stress the superconducting state, turning it first into an Abrikosov lattice of vortices, and eventually destroying the condensate altogether. On the other hand, for sufficiently strong magnetic fields and low-to-moderate isospin chemical potential, the ground state of quantum chromodynamics (QCD) is expected to be a spatially modulated condensate of neutral pions, induced by the chiral anomaly: the chiral soliton lattice (CSL). We map the phase diagram of QCD as a function of isospin chemical potential and magnetic field in the part of the parameter space accessible to a low-energy effective field theory description of QCD. Our main result is an explicit account of the competition between the CSL and the Abrikosov vortex lattice. This is accomplished by adopting a fast numerical algorithm for finding the vortex lattice solution of the equation of motion and the corresponding Gibbs energy. We find that the Abrikosov vortex lattice phase persists in the phase diagram, separating the uniform charged pion BEC phase from the CSL phase. The precise layout of the phase diagram depends sensitively on the choice of the vacuum pion mass.

## 1 Introduction

Understanding the phase diagram of quantum chromodynamics (QCD) remains one of the major unresolved problems in particle physics. The slow pace of progress towards this goal is largely due to the infamous sign problem, which lattice Monte Carlo simulations of QCD suffer from at nonzero baryon chemical potential, $\mu_B$. This conundrum has triggered interest in alternative theories of dense quark matter where the sign problem is absent.

One of such alternative theories is QCD with two light quark flavors and nonzero *isospin* chemical potential, $\mu_I$ [1]. We will refer to this theory as isospin QCD (iQCD). In iQCD, low-density matter is realized by bosons rather than fermionic nucleons. When $\mu_I$ exceeds the vacuum pion mass, $m_\pi$, charged pions undergo Bose–Einstein condensation (BEC). Since the energy scale at which the phase transition to the BEC phase occurs is controlled by the pion mass, it is possible to analyze the properties of such dense matter using the low-energy effective field theory of QCD: the chiral perturbation theory (ChPT) [2]. Together with the absence of sign problem in iQCD, this opens the possibility to confront precision analytic computations within ChPT [3–5] with state-of-the-art lattice simulations of iQCD [6].

The effect of strong magnetic fields on quark matter [7,8] is of great phenomenological interest due to its relevance for heavy-ion collisions and the astrophysics of neutron stars. In particular, the magnetism of quark matter exhibits intriguing properties arising from the chiral anomaly of QCD [9]. In Ref. [10] it was shown that at nonzero $\mu_B$ and in a sufficiently strong magnetic field, the ground state of QCD carries a spatially modulated condensate of neutral pions, dubbed chiral soliton lattice (CSL). This novel state of matter arises from the anomalous coupling of neutral pions to the electromagnetic field, and carries a topologically generated baryon number. Similar topological crystalline phases were subsequently predicted to occur in dense QCD under rotation [11–13] and in QCD-like theories with quarks in a real or pseudoreal representation of the color gauge group [14].

The aim of the present paper is to investigate the thermodynamics of iQCD in an external magnetic field (and at zero temperature). While the presence of the magnetic field reinstates the sign problem, the physical properties of iQCD in external magnetic fields make this goal worth pursuing on its own. Namely, the pion BEC carries electric charge and as such behaves as an electromagnetic superconductor. Weak magnetic fields should be entirely expelled from such a medium by the Meissner effect. On the other hand, sufficiently strong magnetic fields should destroy the condensate. In so-called type-II superconductors, the transition between the uniform BEC phase and the normal state without BEC proceeds in two steps as the magnetic field is cranked up. In the intermediate stage, the magnetic flux penetrates the superconducting medium through a periodic array of vortices, called the Abrikosov lattice. A first sketch of the phase diagram of the pion superfluid as a function of $\mu_I$ and magnetic field appeared nearly a half-century ago [15], including an explicit solution for a single isolated vortex [16]. These early analyses were recently revisited and improved using the model independent approach of ChPT [17]. The Abrikosov vortex lattice in iQCD in external magnetic fields was considered in Ref. [18].

[a] e-mail: martin.gronli96@gmail.com (corresponding author)
[b] e-mail: tomas.brauner@uis.no

The present paper improves upon the existing literature in two main aspects. First, the chiral anomaly affects the magnetism of iQCD in the same way as that of QCD with nonzero $\mu_B$. Hence, sufficiently strong magnetic fields will inevitably lead to the formation of CSL. It is therefore mandatory to consider the competition of charged pion BEC and neutral pion CSL to have a physically adequate picture of the phase diagram of iQCD. Yet, the possible existence of CSL in iQCD has not been investigated before.[1] Without a detailed analysis, it is not even clear whether the phase carrying the Abrikosov lattice of charged pion vortices will survive in the phase diagram.

Second, the analysis of the vortex lattice in Ref. [18] was carried out using a semi-analytic approximation, valid near the upper critical field where the back-reaction of the pion condensate to the external magnetic field is very weak. In this paper, we provide a numerical solution for the vortex lattice that is complete within the tree level of ChPT without further approximations. We do so by adapting a fast momentum-space numerical algorithm for finding vortex lattice solutions in the Ginzburg–Landau theory, known in condensed-matter physics [20,21]. An alternative numerical approach, based on coordinate space discretization, was put forward recently in Ref. [22].

The plan of the paper is as follows. Section 2 reviews the basic setup of ChPT for two light quark flavors including the effects of nonzero isospin chemical potential and magnetic field. We take this opportunity to summarize at one place the technical assumptions that the following sections base upon. A brief reminder of the thermodynamics of physical systems in external magnetic fields is also included. In Sect. 3, we review the physics of uniform charged pion BEC and calculate its Gibbs energy in the external field. This serves as a starting point for finding the thermodynamic equilibrium by comparing several candidate ground states. In Sect. 4 we do the same for the neutral pion CSL. While trading $\mu_B$ for $\mu_I$ as compared to Ref. [10] is next-to-trivial, a novel element here is that we likewise treat the magnetic field as dynamical and pin down the CSL state by minimizing the Gibbs energy. This stands in contrast to all previous accounts of the CSL in quark matter, where the magnetic field was treated as a fixed background. Section 5 is the core of the paper. Here we describe in detail the algorithm for finding a self-consistent vortex lattice solution to the equations of motion. Finally, in Sect. 6 we put all the bits together and map the phase diagram of iQCD as a function of isospin chemical potential and magnetic field. The concluding Sect. 7 is reserved for additional comments.

Appendix A discusses the effect of electrostatic fields arising from the electric charge carried by the BEC and CSL states. While a strict thermodynamic limit may then not exist, it is shown that one can choose the system volume so that our analysis is not invalidated by finite-volume corrections and at the same time the effect of electric fields can be neglected. In Appendix B, we return to the question whether iQCD in an external magnetic field actually is a type-I or a type-II superconductor. We derive a sufficient condition for the existence of a phase with an inhomogeneous pion condensate in the phase diagram, and show that this condition is satisfied for the physical pion mass.

## 2 Effective field theory for QCD at nonzero isospin chemical potential and magnetic field

The analysis presented in this paper is based on the following schematic Lagrangian,

$$\mathcal{L} = \mathcal{L}_{\text{QED}} + \mathcal{L}_{\text{ChPT}} + \mathcal{L}_{\text{WZW}}. \qquad (1)$$

The first contribution is the usual free Maxwell Lagrangian with a gauge-fixing term if needed,

$$\mathcal{L}_{\text{QED}} = -\frac{1}{4} F_{\mu\nu} F^{\mu\nu} + \mathcal{L}_{\text{g.f.}}, \qquad (2)$$

with $F_{\mu\nu} \equiv \partial_\mu A_\nu - \partial_\nu A_\mu$. Moreover,

$$\mathcal{L}_{\text{ChPT}} = \frac{f_\pi^2}{4} \big[ \text{Tr}(D_\mu \Sigma^\dagger D^\mu \Sigma) + m_\pi^2 \, \text{Tr}(\Sigma + \Sigma^\dagger) \big] \qquad (3)$$

is the leading-order Lagrangian of ChPT for two light quark flavors. This depends on two empirically determined parameters, the pion decay constant $f_\pi$ and the vacuum pion mass $m_\pi$. The matrix $\Sigma \in \text{SU}(2)$ encodes the three pion degrees of freedom. Its covariant derivative $D_\mu \Sigma$ couples it to gauge fields associated with the chiral symmetry of QCD. In presence of the electromagnetic gauge potential $A_\mu$, the covariant derivative reads

$$D_\mu \Sigma \equiv \partial_\mu \Sigma - \frac{\mathrm{i}}{2} e A_\mu [\tau_3, \Sigma], \tag{4}$$

where $\tau_3$ is the third Pauli matrix and $e$ is the electromagnetic coupling. A nonzero isospin chemical potential $\mu_\mathrm{I}$ is implemented as a constant temporal background, $A_0 = \mu_\mathrm{I}/e$.

---
[1] The only exception we are aware of is Ref. [19] which, however, only considered a limiting case of CSL, the neutral pion domain wall.

Finally, the last term in Eq. (1) is the anomalous Wess–Zumino–Witten (WZW) term. The full WZW term depends on all three pion fields in a rather complicated manner which we shall not reproduce here; see e.g. Ref. [9]. We will only need a reduced version of the WZW term obtained by discarding the charged pion degrees of freedom and only keeping the neutral pion field, $\pi^0$. This reduced WZW term reads

$$\mathcal{L}_\mathrm{WZW} = \frac{e\mu_\mathrm{I}}{8\pi^2} \mathbf{B} \cdot \boldsymbol{\nabla}\phi, \tag{5}$$

where $\phi \equiv \pi^0/f_\pi$ is a dimensionless neutral pion field. This differs from the WZW term at nonzero $\mu_\mathrm{B}$ used in Ref. [10] to study the CSL in QCD only by a factor of two [23].

It has been proposed that augmenting QCD with dynamical electromagnetic fields leads to a modification of the ChPT Lagrangian by a term that explicitly breaks the isospin SU(2) symmetry [24]. This is responsible for the electromagnetic splitting of pion masses already at tree level. Such a correction to the effective Lagrangian is however numerically very small and we neglect it here for simplicity.

### 2.1 Thermodynamics of iQCD in external magnetic fields

In this paper, we will address the equilibrium properties of iQCD in an external magnetic field. This means that we will assume the magnetic field, and hence the potential $A_\mu$, to be static. We will also assume the absence of any electric field. The latter can be ensured by making the iQCD matter locally electrically neutral in equilibrium. This is a nontrivial assumption since both the charged pion BEC and the CSL carry electric charge. Other degrees of freedom are therefore needed to maintain local charge neutrality. Alternatively, one may argue that, depending on the volume of the system, the effect of electric fields can be neglected even in absence of additional degrees of freedom. This is justified in detail in Appendix A.

With the above assumptions, the equilibrium free energy density of iQCD in the region of the parameter space accessible by ChPT can be written as

$$\mathcal{F} = \frac{\mathbf{B}^2}{2} + \mathcal{H}_\mathrm{ChPT} + \mathcal{H}_\mathrm{WZW}, \tag{6}$$

where $\mathcal{H}$ indicates the canonical Hamiltonian density and the various contributions match those to the Lagrangian (1). The free energy is a local functional of the pion fields and the magnetic field, $\mathbf{B} = \boldsymbol{\nabla} \times \mathbf{A}$. This field is however affected by the back-reaction of the iQCD matter. In order to correctly account for the boundary conditions imposed by the external magnetic field, we have to Legendre-transform to the Gibbs energy density,

$$\mathcal{G} \equiv \mathcal{F} - \mathbf{B} \cdot \mathbf{H}, \tag{7}$$

where $\mathbf{H} = \mathbf{B} - \mathbf{M}$ and $\mathbf{M}$ is the magnetization of the iQCD matter in the external field. In order to determine the thermodynamic equilibrium, we therefore have to minimize the spatial integral of

$$\mathcal{G} = \frac{\mathbf{B}^2}{2} - \mathbf{B} \cdot \mathbf{H} + \mathcal{H}_\mathrm{ChPT} + \mathcal{H}_\mathrm{WZW} \tag{8}$$

as a functional of $\mathbf{A}$ and the pion fields. The external field $\mathbf{H}$ is fixed, and assumed to be uniform throughout this paper.

In practice, it does not seem feasible to find the absolute minimum of the Gibbs energy directly. What we will rather do is to find several specific solutions to the classical equations of motion and evaluate their respective Gibbs energies. The phase diagram is then mapped by comparing the Gibbs energies of these candidate equilibrium states. The following candidate states will be included in the analysis:

– The vacuum state, corresponding to $\Sigma = \mathbb{1}$.
– The uniform charged pion BEC state, analyzed in Sect. 3.
– The neutral pion CSL state, investigated in Sect. 4.
– The Abrikosov vortex lattice state, explored in Sect. 5.

### 2.2 Power counting

An effective field theory such as ChPT relies on a power-counting scheme to be predictive. Within ChPT (see for instance Ref. [25]) it is usual to count each derivative as well as the pion mass as $\mathcal{O}(p^1)$. Including the kinetic term for the electromagnetic field in the Lagrangian (1) can be made consistent with this power-counting if we count $A_\mu = \mathcal{O}(p^0)$ and $e = \mathcal{O}(p^1)$. Likewise, $\mu_\mathrm{I} = \mathcal{O}(p^1)$. Then, both $\mathcal{L}_\mathrm{QED}$ and $\mathcal{L}_\mathrm{ChPT}$ in Eq. (1) enter at the same counting order, $\mathcal{O}(p^2)$. This makes it among others possible to consistently analyze vortices in the pion BEC phase of iQCD within ChPT.[2]

As to the CSL phase, an explicit power-counting scheme that applies to nonzero $\mu_\mathrm{B}$ was proposed recently in Ref. [26]. However, this power-counting scheme treats the chemical potential as a large quantity rather than a small perturbation. Unfortunately, it appears impossible to define a power-counting in which all three parts of the Lagrangian (1) would appear at the same order.

This serves as a warning to the reader that the results we report in this paper should strictly speaking not be understood as predictions of an effective field theory of iQCD with a well-defined derivative expansion. We can however still treat Eq. (1) as a model within which we can map the phase diagram of iQCD. Our model is defined as the minimal chiral Lagrangian that captures all the physics needed to address the competition of charged pion condensation and anomaly-induced neutral pion condensation.

---

[2] We are grateful to Naoki Yamamoto for pointing this out to us.

## 3 Uniform charged pion condensate

In the vacuum state, $\Sigma = \mathbb{1}$, the $\mathscr{H}_{\text{ChPT}}$ term in Eq. (8) reduces to $-m_\pi^2 f_\pi^2$, whereas $\mathscr{H}_{\text{WZW}}$ vanishes. Minimizing the Gibbs energy (8) with respect to the dynamical vector potential $\mathbf{A}$ then gives (not surprisingly) $\mathbf{B} = \mathbf{H}$, and subsequently

$$\mathscr{G}_{\text{vac}} = -\frac{\mathbf{H}^2}{2} - m_\pi^2 f_\pi^2. \tag{9}$$

For $\mu_I > m_\pi$, another uniform stationary state, carrying a BEC of charged pions, exists. This state is a superconductor exhibiting Meissner effect, that is $\mathbf{B} = \mathbf{0}$. This corresponds to the electromagnetic potential $A_\mu = (\mu_I/e, \mathbf{0})$. The pion condensate is parameterized by [1]

$$\Sigma = \mathbb{1}\cos\alpha + i\tau_2 \sin\alpha, \quad \cos\alpha = \frac{m_\pi^2}{\mu_I^2} = \frac{1}{x^2}, \tag{10}$$

where $x \equiv \mu_I/m_\pi$ is a dimensionless isospin chemical potential. This result is by now so well-known that we skip the details of its derivation; see Ref. [27] for a pedagogical account. A simple algebra then leads to an expression for the Gibbs energy density of the uniform pion BEC state,

$$\mathscr{G}_{\text{BEC}} = -\frac{1}{2}m_\pi^2 f_\pi^2 \left(x^2 + \frac{1}{x^2}\right). \tag{11}$$

In the following, we will use the vacuum Gibbs energy (9) as a reference and relate the Gibbs energies of the other candidate states to it. Thus, the relative Gibbs energy density of the uniform charged pion BEC is

$$\boxed{\Delta\mathscr{G}_{\text{BEC}} \equiv \mathscr{G}_{\text{BEC}} - \mathscr{G}_{\text{vac}} = \frac{\mathbf{H}^2}{2} - \frac{1}{2}m_\pi^2 f_\pi^2 \left(x - \frac{1}{x}\right)^2.} \tag{12}$$

Should we assume that the charged pion BEC is a type-I superconductor, we could extract from Eq. (12) the critical magnetic field that destroys the superconducting order,

$$H_c = m_\pi f_\pi \left(x - \frac{1}{x}\right) = f_\pi \left(\mu_I - \frac{m_\pi^2}{\mu_I}\right), \tag{13}$$

in agreement with Ref. [17]. Above this field, the uniform charged pion BEC is thermodynamically disfavored by the energy cost associated with expelling the magnetic field from its bulk. It however turns out that the charged pion BEC is, in fact, a type-II superconductor in a large part of the parameter space of iQCD. Cranking up the external magnetic field does not destroy the order at once, but leads first to an intermediate phase where the condensate is penetrated by a lattice of vortices carrying magnetic flux. This phase is limited to the magnetic field range $H_{c1} < H < H_{c2}$, where $H_{c1}$ and $H_{c2}$ are known respectively as the lower and upper critical field. Below $H_{c1}$ the uniform BEC state prevails, whereas above $H_{c2}$ the vacuum state is favored over any state containing a condensate of charged pions.

It follows by definition of the various critical fields that $H_{c1} < H_c < H_{c2}$. Until we actually determine the lower and upper critical fields, Eq. (13) will therefore still serve as a useful reference, indicating the scale of magnetic field that severely affects the uniform BEC state.

## 4 Chiral soliton lattice

The CSL state carries a spatially inhomogeneous condensate of neutral pions. We can therefore restrict the SU(2)-valued matrix field $\Sigma$ to the ansatz $\Sigma = e^{i\tau_3 \phi}$, where $\phi$ is the dimensionless neutral pion field, introduced in Eq. (5). Upon inserting this ansatz in the Lagrangian (1) and using the result (9) for the Gibbs energy of the vacuum, we can bring the relative Gibbs energy density of neutral pions to the form

$$\Delta\mathscr{G}_{\text{CSL}} = \frac{1}{2}(\mathbf{B} - \mathbf{H})^2 + \frac{f_\pi^2}{2}\left[(\boldsymbol{\nabla}\phi)^2 + 2m_\pi^2(1 - \cos\phi)\right] - \frac{e\mu_I}{8\pi^2}\mathbf{B} \cdot \boldsymbol{\nabla}\phi. \tag{14}$$

Keeping in mind that this should be treated as a local functional of $\phi$ and $\mathbf{A}$, the stationarity condition for the magnetic field becomes $\boldsymbol{\nabla} \times \mathbf{B} = \mathbf{0}$. In order to simplify the analysis, we will assume that the magnetic field $\mathbf{B}$ only varies in the direction of $\mathbf{H}$, which we take by choice to be oriented along the $z$-axis. Together with the zero-divergence constraint on $\mathbf{B}$, this automatically implies that in equilibrium, $\mathbf{B}$ must be a constant vector. Moreover, the $-\mathbf{B} \cdot \mathbf{H}$ term in the Gibbs energy forces $\mathbf{B}$ to be parallel to $\mathbf{H}$.

The partial derivatives of $\phi$ in the transverse directions $x, y$ only enter Eq. (14) squared with a positive sign. It follows that in the equilibrium, $\phi$ can only be modulated in the $z$-direction. The corresponding one-dimensional stationarity condition for $\phi$ reads

$$\partial_z^2 \phi = m_\pi^2 \sin\phi. \tag{15}$$

This is the equation of motion of a simple pendulum, whose closed-form solutions are, up to an overall translation [10],

$$\cos\frac{\phi(zm_\pi/k)}{2} = \operatorname{sn}(zm_\pi/k, k), \tag{16}$$

where sn is one of the Jacobi elliptic functions and $k$, constrained to the range $0 \leq k \leq 1$, the so-called elliptic modulus. The condensate $\Sigma$ defined by this solution exhibits periodic modulation with the period

$$\ell = \frac{2kK(k)}{m_\pi}, \tag{17}$$

where $K(k)$ is the complete elliptic integral of the first kind.

Plugging the solution (16) back into Eq. (14), we get the *spatially averaged* Gibbs energy density of the CSL state as a function of $\mathbf{B}$ and $k$,

$$\Delta \bar{\mathscr{G}}_{\text{CSL}} = \frac{1}{2}(B-H)^2 + 2m_\pi^2 f_\pi^2 \left[ 1 - \frac{1}{k^2} + \frac{2}{k^2} \frac{E(k)}{K(k)} \right] - \frac{e\mu_I m_\pi B}{8\pi k K(k)}, \tag{18}$$

where $E(k)$ is the complete elliptic integral of the second kind. This is a quadratic polynomial in $B$, which gives at once the solution for the magnetic field $\mathbf{B}$ in terms of $\mathbf{H}$,

$$B = H + \frac{e\mu_I m_\pi}{8\pi k K(k)}. \tag{19}$$

Inserting this back in Eq. (18) leads finally to an expression for the Gibbs energy density of the CSL state in terms of a sole free parameter $k$,

$$\boxed{\begin{aligned}\Delta \bar{\mathscr{G}}_{\text{CSL}} = \frac{H^2}{2} &- \frac{1}{2}\left[ H + \frac{e\mu_I m_\pi}{8\pi k K(k)} \right]^2 \\ &+ 2m_\pi^2 f_\pi^2 \left[ 1 - \frac{1}{k^2} + \frac{2}{k^2}\frac{E(k)}{K(k)} \right].\end{aligned}} \tag{20}$$

In order to find the actual equilibrium CSL state, this is to be minimized with respect to $k$, which leads to the condition

$$\frac{E(k)}{k} = \frac{e\mu_I}{32\pi m_\pi f_\pi^2}\left[ H + \frac{e\mu_I m_\pi}{8\pi k K(k)} \right]. \tag{21}$$

This has to be solved numerically.

### 4.1 Chiral limit

Since the analytic expressions for the CSL state found above are rather involved, it is instructive to inspect their behavior in the limiting case of vanishing vacuum pion mass, i.e. the chiral limit, $m_\pi \to 0$. This can be done either by making use of the asymptotic properties of the elliptic integrals, or by setting $m_\pi = 0$ and minimizing the Gibbs energy (14) directly. While the latter approach is much easier, we choose the former, which also covers small but nonzero $m_\pi$.

It follows from Eq. (21) that for very small $m_\pi$, the optimal value of $k$ is very small as well. By approximating both complete elliptic integrals with the leading term of their Taylor series, $E(k) \approx K(k) = \pi/2 + \mathcal{O}(k^2)$, Eq. (21) reduces to

$$\frac{k}{m_\pi} \approx \frac{2f_\pi}{\Omega H}(1-\Omega^2), \quad \Omega \equiv \frac{e\mu_I}{8\pi^2 f_\pi}, \tag{22}$$

where $\Omega$ is a dimensionless measure of the magnitude of the WZW term. Inserting this in Eqs. (17) and (19), we get the corresponding values for the period of the CSL solution and the dynamical magnetic field,

$$\ell \approx \frac{2\pi f_\pi}{\Omega H}(1-\Omega^2), \quad B \approx \frac{H}{1-\Omega^2}. \tag{23}$$

Finally, the relative Gibbs energy density of the CSL state near the chiral limit follows from Eq. (20),

$$\Delta \bar{\mathscr{G}}_{\text{CSL}} \approx -\frac{\Omega^2 H^2}{2(1-\Omega^2)}. \tag{24}$$

Except for the asymptotic expression (22) for $k/m_\pi$, all the above results for the chiral limit can also be derived by setting $m_\pi = 0$ in Eq. (14), upon which the Gibbs energy density only depends on $\mathbf{B}$ and the gradient of $\phi$ and its direct minimization is trivial. For the record, we remark that the corresponding expression for the Gibbs energy density of the uniform charged pion BEC in the chiral limit follows from Eq. (12),

$$\Delta \mathscr{G}_{\text{BEC}} = \frac{1}{2}(\mathbf{H}^2 - f_\pi^2 \mu_I^2). \tag{25}$$

### 4.2 Comparison to CSL at fixed magnetic field

As already briefly mentioned in the introduction, previous treatments of the CSL in QCD were based on the setup where the external magnetic field is a fixed background, which amounts to setting $\mathbf{B} = \mathbf{H}$. This raises a natural question to what extent the approach used here, where $\mathbf{B}$ is a dynamical field that is self-consistently solved for, is more accurate.

Upon replacing $\mu_I$ with $2\mu_B$, previously published results for CSL in QCD [10] are recovered by neglecting the second term on the right-hand side of Eq. (19) as well as the terms in Eqs. (20) and (21), proportional to $e^2$. The latter capture the back-reaction of the neutral pions to the magnetic field, which is absent if $\mathbf{B}$ is treated as fixed.

The validity of this approximation relies on a single condition that should be satisfied in equilibrium,

$$\frac{e\mu_I m_\pi}{8\pi k K(k)} \ll H. \tag{26}$$

Given that $E(k)/k$ is monotonously decreasing on $(0, 1]$ and bounded from below by one, it follows from Eq. (21) that within the same approximation, the CSL state can only exist in equilibrium above the critical field

$$H_{\text{CSL}} = \frac{32\pi m_\pi f_\pi^2}{e\mu_I}. \tag{27}$$

For $k$ that is not too small, this automatically guarantees that the condition (26) holds, as long as the parameter $\Omega$ defined by Eq. (22) is much smaller than one. For very small $k$, we can use the approximate expression (22) to arrive at the same condition, $\Omega \ll 1$. This is in turn satisfied if $\mu_I$ is much smaller than the cutoff scale of ChPT, $4\pi f_\pi$, which is something we expect from a quantity of order $\mathcal{O}(p^1)$.

We conclude that in practice, the corrections to the equilibrium properties of the CSL, caused by treating $\mathbf{B}$ as a dynamical field, are numerically negligible. This can be seen as an a posteriori justification of the assumption made in previous works that the external magnetic field is a fixed, non-dynamical background. That is no longer true for the superconducting charged pion BEC phase, where the dynamical nature of $\mathbf{B}$ is essential for getting the basic physics right.

## 5 Abrikosov vortex lattice

In this section, we will address in detail the possible appearance of a phase with an inhomogeneous charged pion condensate supporting a lattice of magnetic vortices. Our approach follows closely that of Refs. [20,21]. The basic idea is to rewrite the free energy of the magnetic and pion fields in terms of manifestly gauge-invariant variables. Assuming spatial periodicity of the equilibrium state in the plane transverse to the external field $\mathbf{H}$, the various fields are then expanded in a Fourier series; the specific ansatz we use is described in Sect. 5.2. Next, the classical equations of motion are cast in a form that streamlines iterative solution for the Fourier coefficients; this is detailed in Sect. 5.3. This procedure makes it possible to solve numerically the equations of motion at fixed average magnetic flux density, $\bar{\mathbf{B}}$. The latter is related to the external field $\mathbf{H}$ using a version of the virial theorem, derived in Ref. [28]; see Sect. 5.4. Having at hand both the free energy and the external field finally allows us to calculate the Gibbs energy of the vortex lattice.

Our analysis is based on a few simplifying assumptions. These do not constitute any approximation to our model (1) per se, but rather help to select a class of relevant solutions to the equations of motion. First, we assume the absence of any neutral pion condensate. Hence, the $SU(2)$-valued matrix field $\Sigma$ takes the generic form

$$\Sigma = \frac{1}{f_\pi}(\sigma \mathbb{1} + i\tau_1 \pi_1 + i\tau_2 \pi_2), \quad (28)$$

where $\pi_{1,2}$ are the real scalar fields representing charged pions and $\sigma$ is fixed implicitly by the normalization condition $\sigma^2 + \pi_1^2 + \pi_2^2 = f_\pi^2$. It will be convenient to group $\pi_{1,2}$ into a single complex field,

$$\pi \equiv \pi_1 + i\pi_2. \quad (29)$$

Second, we will be looking for static solutions to the equations of motion, and thus assume the electromagnetic gauge potential $A_\mu$ to take the form $A_\mu = (\mu_I/e, \mathbf{A})$ with a time-independent $\mathbf{A}$. With these assumptions, the free energy density implied by the Lagrangian (1) becomes

$$\mathscr{F} = \frac{\mathbf{B}^2}{2} + \frac{1}{2}(\nabla\sigma)^2 + \frac{1}{2}|(\nabla - ie\mathbf{A})\pi|^2 - \frac{1}{2}\mu_I^2 \pi^*\pi - f_\pi m_\pi^2 \sigma. \quad (30)$$

### 5.1 Abrikosov solution near $H_{c2}$

Let us start with the extreme case of external field close to the upper critical field $H_{c2}$. Taking this limit leads to two simplifications. First, the condensate of charged pions becomes very small and the free energy density (30) can therefore be expanded in powers of $\pi$. Second, the back-reaction of the condensate to the magnetic field is negligible and we can therefore treat $\mathbf{B}$ as fixed, $\mathbf{B} = \mathbf{H}$.

The linearized equation of motion for $\pi$, obtained by expanding the free energy density (30) to second order in $\pi$, reads

$$-(\nabla - ie\mathbf{A})^2 \pi = (\mu_I^2 - m_\pi^2)\pi. \quad (31)$$

Solving this equation is equivalent to the usual Landau level eigenvalue problem. We define the gauge-invariant operator $\Pi \equiv -i\nabla - e\mathbf{A}$, and in terms of its components the annihilation and creation operators,

$$a \equiv \frac{1}{\sqrt{2eH}}(\Pi_x + i\Pi_y), \quad a^\dagger \equiv \frac{1}{\sqrt{2eH}}(\Pi_x - i\Pi_y). \quad (32)$$

Our new operators satisfy the simple commutation relation $[\Pi_x, \Pi_y] = ieH$, and hence $[a, a^\dagger] = 1$. The linearized equation of motion (31) now acquires the form

$$eH(2a^\dagger a + 1)\pi = (\mu_I^2 - m_\pi^2)\pi. \quad (33)$$

Since the eigenvalues of $a^\dagger a$ are quantized in terms of non-negative integers, it follows immediately that a nontrivial solution exists only if $\mu_I^2 - m_\pi^2 \geq eH$. This gives the well-known result for the upper critical field,

$$H_{c2} = \frac{\mu_I^2 - m_\pi^2}{e}. \quad (34)$$

With a little extra effort, we can also extract some details about the profile of the condensate that minimizes the free energy. This must be annihilated by $a$. Following Refs. [20,21], we will now switch to the exponential parameterization for the charged pion field $\pi$,

$$\pi \equiv \sqrt{\omega}\, e^{i\theta}. \quad (35)$$

We will also trade the vector potential $\mathbf{A}$ for the gauge-invariant condensate supervelocity,

$$\mathbf{Q} \equiv \mathbf{A} - \frac{\nabla \theta}{e}. \tag{36}$$

Considering separately the real and imaginary parts of the condition $(\Pi_x + i\Pi_y)\sqrt{\omega}\, e^{i\theta} = 0$ then leads to a relation between $\mathbf{Q}$ and the gradient of $\omega$, which can be put in a neat vector form,

$$\mathbf{Q} = \frac{1}{2e} \frac{\nabla \omega \times \hat{z}}{\omega}. \tag{37}$$

This relates the supervelocity of the condensate to the spatial variation of its magnitude in a way that is independent of the choice of gauge for $\mathbf{A}$.

Equation (37) will be used as a benchmark below when we search for the vortex lattice solution away from the upper critical field. Further details about Abrikosov's analytic solution for the vortex lattice near $H_{c2}$, tailored to iQCD, can be found in Ref. [18].

### 5.2 Vortex lattice ansatz away from $H_{c2}$

With the exponential parameterization (35) and the definition (36) at hand, the free energy density (30) can be written solely in terms of the gauge-invariant variables $\omega$ and $\mathbf{Q}$,

$$\mathscr{F} = \frac{\mathbf{B}^2}{2} + \frac{(\nabla\omega)^2}{8}\left(\frac{1}{f_\pi^2 - \omega} + \frac{1}{\omega}\right) \\ - \frac{1}{2}(\mu_I^2 - e^2\mathbf{Q}^2)\omega - f_\pi m_\pi^2\sqrt{f_\pi^2 - \omega}, \tag{38}$$

where we eliminated $\sigma$ in terms of $\omega$ by using the constraint $\sigma^2 + \omega = f_\pi^2$.

We will search for stationary states of the free energy, obtained by integrating Eq. (38) over space, that carry a periodic lattice of magnetic vortices, aligned with the external field $\mathbf{H}$. We will assume that such states are strictly two-dimensional in that both $\omega$ and $\mathbf{B}$ only depend on the transverse coordinates $x, y$. Moreover, we will assume that $\mathbf{B}$ is parallel to $\mathbf{H}$ everywhere, that is, $\mathbf{B} = (0, 0, B(x, y))$. In terms of the two-dimensional position vector $\mathbf{r} = (x, y)$, the magnitude of the condensate and the magnetic field can then be Fourier-expanded as

$$\omega(\mathbf{r}) = \sum_{\mathbf{K}} a_{\mathbf{K}}(1 - \cos \mathbf{K} \cdot \mathbf{r}),$$
$$B(\mathbf{r}) = \bar{B} + \sum_{\mathbf{K}} b_{\mathbf{K}} \cos \mathbf{K} \cdot \mathbf{r}, \tag{39}$$

where $\mathbf{K}$ are vectors from the reciprocal lattice; only $\mathbf{K} \neq \mathbf{0}$ are included in the sum. Also, $\bar{B}$ is the spatial average of the magnetic flux density.

The direct lattice in real space, corresponding to points $\mathbf{R}$ such that $\mathbf{K} \cdot \mathbf{R}$ is an integer multiple of $2\pi$ for all $\mathbf{K}$, carries the cores of vortices at which the charged pion condensate vanishes. For a generic lattice, the direct and reciprocal lattice vector can be parameterized by two integers $m, n$ as

$$\mathbf{R}_{mn} \equiv (mx_1 + nx_2, ny_2),$$
$$\mathbf{K}_{mn} \equiv \frac{2\pi}{S}(my_2, -mx_2 + nx_1), \tag{40}$$

where the lattice basis in real space has been chosen without loss of generality as $(x_1, 0)$ and $(x_2, y_2)$, and $S \equiv x_1 y_2$ is the area of the unit cell of the direct lattice. We will restrict our analysis to a hexagonal lattice ansatz, for which we can choose the parameters of the lattice basis as

$$x_2 = \frac{x_1}{2}, \qquad y_2 = \frac{x_1\sqrt{3}}{2}. \tag{41}$$

This lattice geometry is known to be energetically favored on a fairly general ground [29].

To be able to evaluate the free energy for the ansatz (39), we need to relate the supervelocity $\mathbf{Q}$ to the magnetic field $\mathbf{B}$. To that end, note that near the vortex core at $\mathbf{R} = \mathbf{0}$, the phase of the condensate behaves as $\theta(\mathbf{r}) \simeq n\varphi$, where $\varphi$ is the polar angle and $n$ the winding number of the vortex. This implies that $\nabla\theta(\mathbf{r}) \simeq n\hat{\varphi}/r$. Using the Stokes theorem, we then find that

$$\nabla \times \nabla\theta(\mathbf{r}) = 2\pi n \hat{z}\delta(\mathbf{r}), \tag{42}$$

where the two-dimensional $\delta$-function is supported on the vortex core, corresponding to the line $\mathbf{r} = \mathbf{0}$. In the following, we only consider vortices with a unit winding number, which are usually energetically favored. Taking into account the whole lattice of vortices, the definition (36) then leads to

$$\nabla \times \mathbf{Q}(\mathbf{r}) = \mathbf{B}(\mathbf{r}) - \Phi_0 \hat{z} \sum_{\mathbf{R}} \delta(\mathbf{r} - \mathbf{R}), \tag{43}$$

where $\Phi_0 \equiv 2\pi/e$ is the quantum of magnetic flux. Requiring accordingly that each unit cell of the vortex lattice carries a single quantum of flux, that is $\bar{B} = \Phi_0/S$, fixes the lattice spacing of the hexagonal lattice (41) in terms of $\bar{B}$,

$$x_1 = \sqrt{\frac{4\pi}{\sqrt{3}\, e\bar{B}}}. \tag{44}$$

Working with a singular field such as $\mathbf{Q}$ is detrimental to the prospect of solving for the vortex lattice numerically. This problem can be bypassed by separating out the singular part of $\mathbf{Q}$. We do so by writing $\mathbf{Q}$ as $\mathbf{Q} = \mathbf{Q}_A + \mathbf{Q}_B$, where $\mathbf{Q}_A$ is the supervelocity (37) of the Abrikosov solution. The latter corresponds to a nearly uniform magnetic field, hence

$$\nabla \times \mathbf{Q}_A(\mathbf{r}) = \bar{B}\hat{z} - \Phi_0 \hat{z} \sum_{\mathbf{R}} \delta(\mathbf{r} - \mathbf{R}). \tag{45}$$

It follows that $\nabla \times \mathbf{Q}_B(\mathbf{r}) = [B(\mathbf{r}) - \bar{B}]\hat{z}$. The two components of $\mathbf{Q}$ thus play very different roles: $\mathbf{Q}_A$ carries information about the average magnetic flux density and the location of vortices, whereas $\mathbf{Q}_B$ encodes smooth variations of the magnetic field around the average. Such a smooth field can be Fourier-expanded, and we start with a generic ansatz,

$$\mathbf{Q}_B(\mathbf{r}) = \sum_{\mathbf{K}} \mathbf{c_K} \sin \mathbf{K} \cdot \mathbf{r}. \tag{46}$$

Comparing this to Eq. (39) results in the condition $\mathbf{K} \times \mathbf{c_K} = b_{\mathbf{K}} \hat{z}$. This determines $\mathbf{c_K}$ up to addition of a multiple of $\mathbf{K}$. In order to remove the ambiguity, we make one last assumption on our stationary state ansatz, that is, $\nabla \cdot \mathbf{Q} = 0$. This is satisfied exactly by the solution (37) near $H_{c2}$. It is also true thanks to continuous rotational invariance for a single isolated vortex, that is near the lower critical field $H_{c1}$, and for any $\mathbf{H}$ near the vortex cores. A detailed justification of why $\nabla \cdot \mathbf{Q}$ can be generally expected to be negligibly small is given in the appendix of Ref. [20]. We now have a unique solution for the Fourier coefficients of $\mathbf{Q}_B$, $\mathbf{c_K} = b_{\mathbf{K}}(\hat{z} \times \mathbf{K})/\mathbf{K}^2$. The full supervelocity then assumes the form

$$\mathbf{Q}(\mathbf{r}) = \frac{1}{2e} \frac{\nabla \omega_A(\mathbf{r}) \times \hat{z}}{\omega_A(\mathbf{r})} + \sum_{\mathbf{K}} b_{\mathbf{K}} \frac{\hat{z} \times \mathbf{K}}{\mathbf{K}^2} \sin \mathbf{K} \cdot \mathbf{r}, \tag{47}$$

which is completely fixed by $\bar{B}$ and the Fourier coefficients $b_{\mathbf{K}}$. We have not given a detailed solution for the condensate magnitude $\omega_A$ of the Abrikosov solution (37). For our purposes, it is however sufficient to know its Fourier components as defined by Eq. (39), see Ref. [30],

$$a^A_{\mathbf{K}_{mn}} = -(-1)^{m+mn+n} \exp\left(-\frac{\mathbf{K}_{mn}^2 S}{8\pi}\right). \tag{48}$$

Equations (39)–(41), (44), (47) and (48) determine our vortex lattice ansatz in terms of the average magnetic flux density $\bar{B}$ and the Fourier coefficients $a_{\mathbf{K}}, b_{\mathbf{K}}$. It remains to be seen for what values of the Fourier coefficients this field configuration is a solution to the classical equations of motion. This is addressed in Sect. 5.3. Finally, in Sect. 5.4, we will show how $\bar{B}$ can be related to the external field $\mathbf{H}$.

### 5.3 Ginzburg–Landau equations for iQCD

Here we will derive a set of equations that constitute a basis for numerical iterative solution for the Fourier coefficients $a_{\mathbf{K}}, b_{\mathbf{K}}$, whose results are presented in Sect. 6. We start with the equation of motion for the magnitude of the condensate, $\omega$. A direct variation of the free energy (38) gives

$$(-\nabla^2 + \xi^2)\omega = \frac{1}{2}\left(\omega - \frac{\omega^2}{f_\pi^2}\right)\left\{4(\mu_I^2 - e^2\mathbf{Q}^2)\right.$$
$$+ (\nabla\omega)^2\left[\frac{1}{(f_\pi^2 - \omega)^2} - \frac{1}{\omega^2}\right]$$
$$\left. - \frac{4f_\pi m_\pi^2}{\sqrt{f_\pi^2 - \omega}}\right\} + \xi^2\omega, \tag{49}$$

where the term $\xi^2\omega$ with an arbitrary coefficient $\xi^2$ was added on both sides to ensure stability and convergence of the itera-tive procedure for solving for the Fourier coefficients. In the next step, we insert the Fourier expansion (39) on the left-hand side, multiply by $\cos \mathbf{K}' \cdot \mathbf{r}$ and take a spatial average, which we will indicate with angular brackets, $\langle \cdots \rangle$. Using the property that $\langle \cos \mathbf{K} \cdot \mathbf{r} \cos \mathbf{K}' \cdot \mathbf{r}\rangle = (1/2)\delta_{\mathbf{K}\mathbf{K}'}$, we arrive at an iterative equation for $a_{\mathbf{K}}$,

$$a_{\mathbf{K}} = -\frac{2}{\mathbf{K}^2 + \xi^2}\bigg\langle\bigg[2\left(\omega - \frac{\omega^2}{f_\pi^2}\right)(\mu_I^2 - e^2\mathbf{Q}^2)$$
$$+ \frac{(\nabla\omega)^2}{2f_\pi^2}\left(\frac{\omega}{f_\pi^2 - \omega} - \frac{f_\pi^2 - \omega}{\omega}\right) \tag{50}$$
$$- \frac{2m_\pi^2\omega}{f_\pi}\sqrt{f_\pi^2 - \omega} + \xi^2\omega\bigg]\cos \mathbf{K} \cdot \mathbf{r}\bigg\rangle.$$

Next we consider the equation of motion for $\mathbf{A}$. Here Eq. (38) gives at once

$$\nabla \times \mathbf{B} = -e^2\omega\mathbf{Q}. \tag{51}$$

Taking a curl of this equation and using the fact that $\mathbf{B}$ has a vanishing divergence, we can bring this to the form[3]

$$(\nabla^2 - e^2\zeta^2)\mathbf{B} = e^2\left[\nabla\omega \times \mathbf{Q} + (\omega - \zeta^2)\mathbf{B}\right]. \tag{52}$$

We have again added an extra term with an arbitrary coefficient $\zeta$ to both sides of the equation to ensure convergence of the iterative procedure. In the next step, we use the Fourier expansion (39) on the left-hand side, multiply with $\cos \mathbf{K}' \cdot \mathbf{r}$ and take a spatial average. This leads to an iterative equation for $b_{\mathbf{K}}$,

$$b_{\mathbf{K}} = -\frac{2e^2}{\mathbf{K}^2 + e^2\zeta^2}\bigg\langle\big[Q_y\partial_x\omega - Q_x\partial_y\omega$$
$$+ (\omega - \zeta^2)B\big]\cos \mathbf{K} \cdot \mathbf{r}\bigg\rangle. \tag{53}$$

Equations (50) and (53) constitute the basis for the iterative solution to the equations of motion. The convergence of the iteration can however be further accelerated by a uniform rescaling of $\omega$ after each application of Eq. (50). This is done by the replacement

$$\omega \to (1 + c)\omega, \tag{54}$$

or equivalently $a_{\mathbf{K}} \to (1 + c)a_{\mathbf{K}}$. The value of the constant $c$ is optimized by requiring that upon the rescaling, the spatial average of the free energy density (38) has a vanishing derivative with respect to $c$. Since we expect $c$ to be small, we can simplify this condition by expanding to first order in $c$. The resulting linear equation for $c$ is solved by

$$c = -\frac{\left\langle \frac{(\nabla\omega)^2}{\omega}\left(\frac{f_\pi^2}{f_\pi^2-\omega}\right)^2 + \frac{4f_\pi m_\pi^2 \omega}{\sqrt{f_\pi^2-\omega}} - 4\omega(\mu_I^2 - e^2\mathbf{Q}^2)\right\rangle}{2\left\langle \frac{f_\pi^4(\nabla\omega)^2}{(f_\pi^2-\omega)^3} + \frac{f_\pi m_\pi^2 \omega^2}{(f_\pi^2-\omega)^{3/2}}\right\rangle}. \tag{55}$$

---

[3] Taking the curl of $\omega\mathbf{Q}$ also produces a term proportional to $\omega\nabla \times \nabla\theta$. This can however be dropped in spite of the singular behavior of $\theta$ near the vortex cores. This is thanks to the presence of the factor $\omega$ which vanishes at the cores.

### 5.4 Virial theorem and the Gibbs energy

So far we have kept the average magnetic flux density $\bar{B}$ as a fixed input parameter. In order to evaluate the Gibbs energy of the Abrikosov vortex lattice, we must however relate it to the imposed external field $\mathbf{H}$. This can be done elegantly using a version of the virial theorem, put forward in Ref. [28].

Let us start by a uniform coordinate dilatation, $\mathbf{r}' = \mathbf{r}/\lambda$, under which the charged pion field transforms as a scalar,

$$\pi'(\mathbf{r}') = \pi(\mathbf{r}), \quad \text{or} \quad \pi'(\mathbf{r}) = \pi(\lambda\mathbf{r}). \tag{56}$$

The vector potential $\mathbf{A}$ is a covariant vector (1-form), and hence transforms differently,

$$\mathbf{A}'(\mathbf{r}') = \lambda\mathbf{A}(\mathbf{r}), \quad \text{or} \quad \mathbf{A}'(\mathbf{r}) = \lambda\mathbf{A}(\lambda\mathbf{r}). \tag{57}$$

Finally, the magnetic field $\mathbf{B}$ is an antisymmetric rank-two covariant tensor (2-form), and thus transforms as

$$\mathbf{B}'(\mathbf{r}') = \lambda^2\mathbf{B}(\mathbf{r}), \quad \text{or} \quad \mathbf{B}'(\mathbf{r}) = \lambda^2\mathbf{B}(\lambda\mathbf{r}). \tag{58}$$

This last rule agrees with the observation and upon the coordinate dilatation, the magnetic flux through the unit cell of the vortex lattice must remain unchanged (because it is quantized in units of $\Phi_0$), whereas the area of the unit cell is rescaled by the factor $1/\lambda^2$.

In thermodynamic equilibrium, the variation of the free energy under the above coordinate rescaling must vanish. This can be ensured by inserting our scaling relations in Eq. (38), taking the spatial average and setting the derivative thereof at $\lambda = 1$ to zero. This leads to the condition

$$\mathbf{H} \cdot \bar{\mathbf{B}} = \left\langle \frac{(\nabla\omega)^2}{8}\left(\frac{1}{f_\pi^2-\omega} + \frac{1}{\omega}\right) + \frac{1}{2}e^2\mathbf{Q}^2\omega + \mathbf{B}^2 \right\rangle, \tag{59}$$

where we have used the thermodynamic relation $\partial\bar{\mathscr{F}}/\partial\bar{\mathbf{B}} = \mathbf{H}$. By combining this with the definition (7) of Gibbs energy and the previously calculated Gibbs energy density (9) of the vacuum state, we obtain our final result for the spatially averaged Gibbs energy density of the Abrikosov vortex lattice, relative to the vacuum,

$$\boxed{\Delta\bar{\mathscr{G}}_{\text{vortex}} = \frac{\mathbf{H}^2}{2} + \left\langle -\frac{\mathbf{B}^2}{2} - \frac{1}{2}\mu_I^2\omega - f_\pi m_\pi^2\left(\sqrt{f_\pi^2-\omega} - f_\pi\right)\right\rangle.} \tag{60}$$

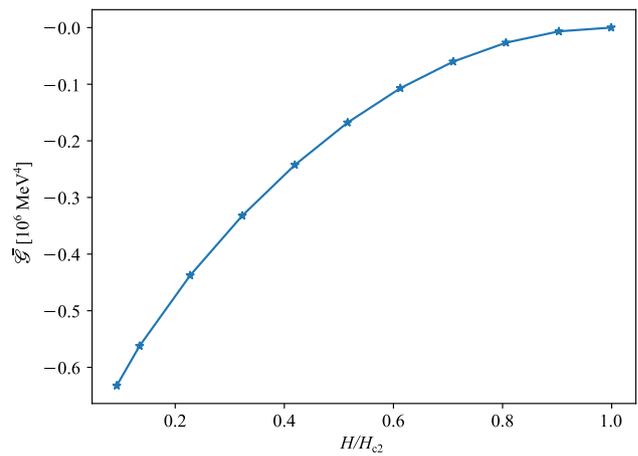

**Fig. 1** Average Gibbs energy density of the Abrikosov vortex lattice, relative to the vacuum state, as a function of $H$ at $\mu_I = 147$ MeV

## 6 Numerical results

All the numerical results presented below were obtained with a fixed value of the pion decay constant, $f_\pi = 92$ MeV. Likewise, we mostly considered the physical pion mass, $m_\pi = 140$ MeV. Sometimes, it is however advantageous to treat the pion mass as a tunable parameter. We use this as a tool to access features of the phase diagram that might otherwise remain hidden. Finally, in the high-energy units that we use throughout the paper, the fine structure constant is $\alpha = e^2/(4\pi)$. Hence the electromagnetic coupling takes the numerical value $e = \sqrt{4\pi\alpha} \approx 0.303$.

### 6.1 Abrikosov vortex lattice

In order to find the Abrikosov vortex lattice solution in equilibrium, we used the following numerical procedure. Initially, we set $b_\mathbf{K} = 0$ and iterated Eqs. (50) and (55) for $a_\mathbf{K}$ a few times. Subsequently, we iterated Eqs. (50), (55) and (53) until convergence was reached. The spatial averages indicated with angular brackets were found by numerical integration over a single unit cell of the direct lattice. In practice, the number of points of the reciprocal lattice included in the calculation was of the order of a thousand.

Before we tackle the problem of finding the phase diagram of iQCD, let us first do some basic checks that the purely numerically obtained vortex lattice solution has the expected physical properties. The upper critical field $H_{c2}$, given analytically by Eq. (34), represents a field at which the vortex lattice solution ceases to be energetically favored over the vacuum state. Hence the relative averaged Gibbs energy density of the vortex lattice (60) should go to zero as $H \to H_{c2}$. This behavior is displayed in Fig. 1. For numerical illustration, we have chosen a value of $\mu_I$ just above the onset of charged pion BEC, $\mu_I = 147$ MeV, for which the upper critical field is small compared to the characteristic scale of QCD, $H_{c2} \approx 6600$ MeV$^2$.

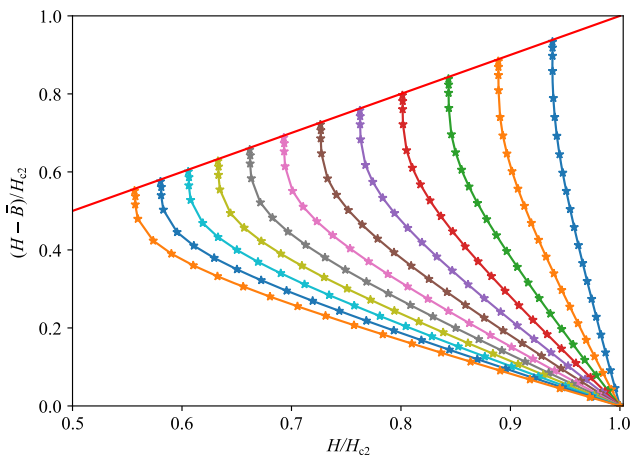

**Fig. 2** Reverse magnetization of the Abrikosov vortex lattice as a function of $H$. To generate these data, the pion mass was tuned to $m_\pi = ef_\pi \approx 27.9$ MeV. The different curves with data points displayed correspond to different values of $\mu_I$ with 1 MeV spacing. The right-most curve corresponds to $\mu_I = 29$ MeV and the left-most one to $\mu_I = 40$ MeV

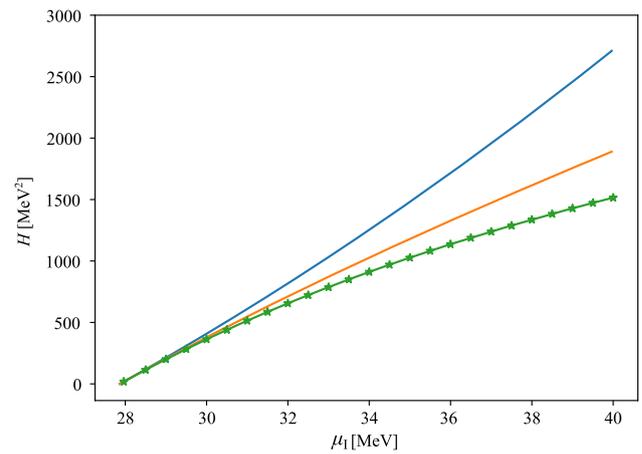

**Fig. 3** Various critical fields for the charged pion condensate. The three curves displayed correspond, from the top to the bottom, to $H_{c2}$, $H_c$ and $H_{c1}$, respectively. To generate these data, the pion mass was tuned to $m_\pi = ef_\pi \approx 27.9$ MeV

The case of the lower critical field $H_{c1}$ is more subtle. In principle, this is defined as the external field at which the Gibbs energy of a single isolated vortex becomes smaller than that of the uniform BEC state [18]. In order to avoid having to deal separately with the isolated vortex solution, we have adopted a different approach. We picture the isolated vortex as an extreme case of a vortex lattice with an infinite lattice spacing. According to Eq. (44), this corresponds to $\bar{B} \to 0$. We take vanishing of $\bar{B}$ as an alternative definition of the lower critical field.

The downside of this approach is that finding the lower critical field through a limit is numerically challenging. The convergence of the search for the critical field turns out to be particularly sensitive to the value of the pion mass. We were not able to obtain sufficient accuracy for the physical pion mass. We shall instead present some numerical results for a lower pion mass, $m_\pi = ef_\pi \approx 27.9$ MeV. This value is not chosen accidentally; it is a threshold value above which the existence of a phase with an inhomogeneous pion condensate can be guaranteed. See Appendix B for a detailed proof of this claim.

It is practically advantageous to focus on the magnetization, $\mathbf{M} = \mathbf{B} - \mathbf{H}$. In Fig. 2, we show the values of magnetization as a function of $H$ for various choices of $\mu_I$; for convenience, the magnetization is displayed with an opposite overall sign. The straight solid line in Fig. 2 corresponds to $\bar{B} = 0$, hence $\bar{M} = -H$. The point of contact of the various data curves with this line defines the lower critical field. Although the numerical approach does not allow us to set $\bar{B} = 0$, we can make a small extrapolation of the data curves in order to identify the points of contact with the straight solid line. The convergence of all the data curves to a single point with $\mathbf{M} = \mathbf{0}$ as $H/H_{c2} \to 1$ is an independent verification of the correct physical behavior of our numerically found solution near the upper critical field.

The values of $H_{c1}$ extracted using this approach are displayed in Fig. 3 as a function of $\mu_I$. We crosschecked these results independently by looking for the value of $H$ where the Gibbs energies of the uniform charged pion BEC and the vortex lattice coincide. For comparison, we also show in the figure the upper critical field $H_{c2}$ (34) and the critical field $H_c$ (13) above which the uniform charged pion BEC state is no longer favored over the vacuum state. As explained in Sect. 3, these three fields are expected to satisfy the hierarchy $H_{c1} < H_c < H_{c2}$, which the numerical results for $H_{c1}$ confirm. All the three fields vanish simultaneously as $\mu_I \to m_\pi$. Below this threshold, no charged pion condensation can occur, uniform or not.

### 6.2 Phase diagram of iQCD

Let us at last address the phase diagram of iQCD, which is the primary target of the present paper. We now know the Gibbs energies of all four candidate states, listed at the end of Sect. 2.1. We take the vacuum state as the reference. The relative Gibbs energies of the uniform charged pion BEC, the neutral pion CSL, and the Abrikosov vortex lattice of charged pions are then given respectively by Eqs. (12), (20) and (60). Our main result is the phase diagram in Fig. 4, obtained using the physical pion mass.

Let us comment on the main qualitative features of the phase diagram. The curve labeled as $H_\text{CSL}$ is defined by setting $\Delta\mathcal{G}_\text{CSL} = 0$. In the region of the phase diagram where charged pion condensation is disfavored by the strong mag-netic field, this describes a phase transition between the vacuum state and the CSL phase. The critical field $H_c$ is shown in Fig. 4 just for orientation. Since for the physical pion mass, we were not able to calculate the lower critical field $H_{c1}$ for the formation of vortex lattice accurately, the $H_c$ curve delimits the part of the phase diagram to which the uniform charged pion BEC phase is localized. The Abrikosov vortex lattice phase occupies a significant part of the phase diagram. The first-order transition between the vortex lattice phase and the CSL phase was pinned down by a numerical comparison of $\Delta \bar{\mathcal{G}}_{CSL}$ and $\Delta \bar{\mathcal{G}}_{vortex}$. This requires the full numerical machinery as laid out in Sect. 5. The position of the triple point at which the vacuum, vortex lattice and CSL phases meet can however be determined simply by solving the condition $H_{CSL} = H_{c2}$. We find the coordinates of the triple point to be $\mu_{I,tr} \approx 505$ MeV and $H_{tr} \approx 0.78$ GeV$^2$. Note that the latter value simultaneously represents the threshold magnetic field for the appearance of the CSL phase in the phase diagram of iQCD.

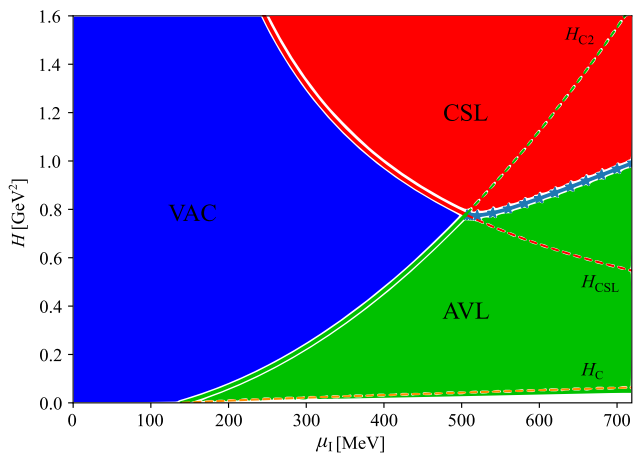

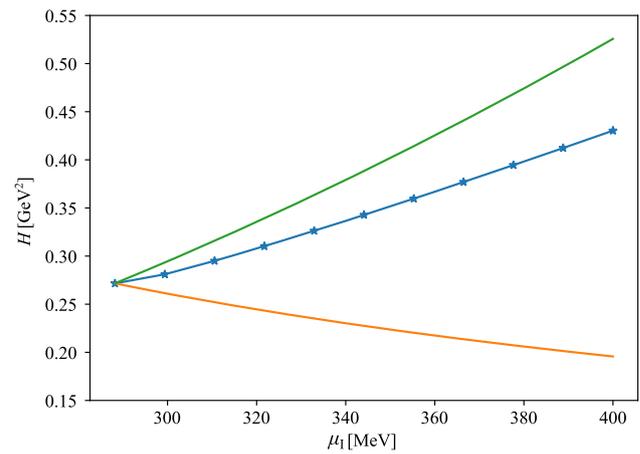

**Fig. 4** Phase diagram of iQCD as a function of external magnetic field and isospin chemical potential for the physical pion mass, $m_\pi = 140$ MeV. The thick solid lines indicate phase transitions between the vacuum (VAC), Abrikosov vortex lattice (AVL) and CSL phases. For orientation, we also show the upper critical field $H_{c2}$ which defines the transition between the VAC and AVL phases, and the curve $H_{CSL}$, defining the transition between the VAC and CSL phases. The lower critical field $H_{c1}$ separating the AVL phase from the uniform BEC phase was not possible to determine accurately due to convergence issues. The phase transition between the BEC and AVL phases however lies below $H_c$, indicated by the thin dashed line near the horizontal axis

**Fig. 5** Competition of the CSL phase and the Abrikosov vortex lattice phase for a low pion mass, $m_\pi = ef_\pi \approx 27.9$ MeV. The top curve indicates the upper critical field $H_{c2}$. The bottom curve corresponds to the field $H_{CSL}$, at which the CSL state becomes favored over the vacuum. The actual phase transition between the vortex lattice and CSL phases is indicated by the middle curve with data points

It is instructive to see how the details of the phase diagram depend on the pion mass, which we to some extent treated as a tunable parameter. First, note that both $H_c$ and $H_{c2}$ only contain $m_\pi$ through the combination $\mu_I^2 - m_\pi^2$. Tuning the pion mass to smaller values and ultimately towards the chiral limit will therefore leave a large part of the phase diagram, for which $\mu_I^2 \gg m_\pi^2$, unaffected. On the other hand, the critical curve $H_{CSL}$ will scale down roughly linearly as $m_\pi$ is lowered; cf. Eq. (27). At the same time, the competition of the vortex lattice and CSL phases to the right of the triple point is affected by changing the pion mass. A detailed numerical comparison shows that for small pion masses, the boundary separating the vortex lattice and CSL phases drifts toward the upper critical field; this trend is clearly visible in Fig. 5. For $m_\pi = ef_\pi$, used in this figure, we find the triple point located at $\mu_{I,tr} \approx 288$ MeV and $H_{tr} \approx 0.27$ GeV$^2$.

In the chiral limit, the critical fields $H_c$ (13) and $H_{c2}$ (34) become equal to each other at $\mu_I = ef_\pi$. The above observations then allow us to draw the following rough picture of the phase diagram of iQCD close to the chiral limit. The vacuum state is only realized in equilibrium for $\mu_I = 0$ and $H = 0$. For $0 < \mu_I < ef_\pi$, the equilibrium state is either the uniform charged pion BEC or the neutral pion CSL. These are separated by a first-order phase transition at $H = H_c = f_\pi \mu_I$. For $ef_\pi < \mu_I$, an intermediate phase supporting the Abrikosov vortex lattice appears. The first-order transition from the vortex lattice phase to the CSL phase appears roughly at $H = H_{c2} = \mu_I^2/e$.

## 7 Summary and discussion

In this paper, we have used ChPT to investigate the phase diagram of QCD in presence of a nonzero isospin chemical potential (iQCD) and external magnetic field. Experience based on available literature on the subject leads to the expectation that the phase diagram includes a phase with a uniform condensate of charged pions, and the anomaly-induced CSL phase with a spatially modulated condensate of neutral pions. Our main goal was to study the interplay of these two orders, focusing specifically on the possibility that another inhomogeneous phase, carrying a lattice of magnetic vortices, appears in an intermediate range of magnetic fields.

While the uniform charged pion BEC state and the CSL state can be treated largely analytically, the Abrikosov vortex lattice requires a fully numerical solution of the classical equations of motion. Here we have improved on previously published results on iQCD by implementing a fast numerical algorithm for finding the vortex lattice solution without any approximations to the underlying ChPT Lagrangian. Our main qualitative result is that for the physical pion mass of $m_\pi = 140$ MeV, the Abrikosov vortex lattice phase indeed appears in the phase diagram, occupying a large range of magnetic fields. Hence iQCD is a type-II superconductor. Type-I superconductivity appears to be possible only for small, unphysical pion masses, $m_\pi < ef_\pi \approx 28$ MeV.

Let us briefly mention one technical aspect of our setup that might be worth pointing out. Namely, in writing down the anomaly-induced WZW term (5), we discarded charged pion degrees of freedom and only kept the neutral pion field, relevant for the CSL state. In hindsight, we have not committed any approximation by doing so. Upon a closer look, it is possible to show that the full WZW term, spelled out for instance in Ref. [9], does not contribute to the classical Gibbs energy of either the uniform charged pion BEC, or the Abrikosov vortex lattice within the two-dimensional ansatz used here.

Finally, let us comment on the fact that the threshold magnetic field for the appearance of the CSL phase in the phase diagram of iQCD was found to be $H_{\text{tr}} \approx 0.78$ GeV$^2$ for the physical pion mass. This is a consequence of the competition between the CSL state and the Abrikosov vortex lattice, as shown in Fig. 4. A natural question arises whether the ChPT setup used in this paper can still be trusted for such strong magnetic fields. On the one hand, it was argued in Ref. [14] that magnetic fields well above the QCD scale do not necessarily preclude the application of a low-energy effective field theory based on the spontaneous breakdown of chiral symmetry. On the other hand, it might then be necessary to take explicitly into account the anisotropy of the QCD vacuum induced by the magnetic field, and in the very least the magnetic field dependence of the input parameters $f_\pi, m_\pi$ of ChPT induced by loop corrections.

We conclude from the above that the existence of the CSL phase in the phase diagram of iQCD for the physical pion mass of $m_\pi = 140$ MeV currently still has the status of a conjecture that requires further investigation. However, the CSL phase certainly is relevant for the thermodynamics of iQCD, since we can treat the pion mass as a tunable parameter and lower it towards the chiral limit. This will reduce the threshold magnetic field for the appearance of the CSL phase to in principle arbitrarily weak fields.

**Acknowledgements** Some of the results presented here appeared previously in the master thesis of one of us (M.S.G.) [31]. The numerical solution of the Abrikosov vortex lattice was done using a PYTHON code, adapted from a MATLAB code, available from the authors of Ref. [32]; see the URL therein. We are indebted to Prabal Adhikari, Jens Oluf Andersen, Andreas Schmitt, Asle Sudbø and Naoki Yamamoto for useful discussions and correspondence. This work has been supported in part by the Grant No. PR-10614 within the ToppForsk-UiS program of the University of Stavanger and the University Fund.

**Data Availability Statement** This manuscript has no associated data or the data will not be deposited. [Authors' comment: The numerical results presented in the paper are based on computations that are outlined in detail in the text, including all the input parameters. This guarantees reproducibility of the results. The computer codes used to generate the numerical results are available from the corresponding author on reasonable request.]

## Appendix A: Effect of electrostatic interaction

It is well-known that in presence of long-range interactions, the thermodynamic limit may be ill-defined as a consequence of the fact that the free energy ceases to be extensive. This applies in particular to systems with nonzero (average) electric charge density, where the thermodynamic limit is invalidated by the non-extensive energy of the ensuing electric field. In ordinary superconductors, the problem is absent thanks to the crystal lattice of positive ions that neutralizes the electric charge of electrons. The situation is however different in the ordered phases of iQCD, discussed in this paper.

One way out is to assume the presence of additional degrees of freedom that neutralize the electric charge carried by the pion condensate. This may actually be appropriate for applications to, for instance, dense matter inside neutron stars. The downside of this approach is that the analysis then necessarily becomes model-dependent. In order to estimate the (non-electrostatic) contribution of such a background to the free energy, one would need additional input.

We will instead take a different route, largely following Ref. [17]. We will show that there is a range of system volumes in which the electrostatic interaction energy in fact can be neglected, even in the absence of additional degrees of freedom. Let us consider a generic system with extensive free energy $F_{\text{ext}}$ due to short-range interactions. If the linear dimension of the system is $R$, this free energy is given parametrically by

$$F_{\text{ext}} \sim \mathscr{F}_{\text{ext}} R^3, \tag{A.1}$$

where $\mathscr{F}_{\text{ext}}$ is the corresponding free energy density. Should the system carry nonzero average electric charge density $\rho$, the long-range Coulomb interaction contributes the energy

$$F_{\text{Coul}} \sim \rho^2 R^5. \tag{A.2}$$

Within iQCD, $\rho$ is related to the isospin density $n_{\text{I}}$ by $\rho = e n_{\text{I}}$. Obviously, the electrostatic interaction dominates in large enough volumes. Its contribution can however be neglected provided the system size is chosen judiciously,

$$R^2 \lesssim \frac{\mathscr{F}_{\text{ext}}}{\rho^2}. \tag{A.3}$$

To get a concrete bound on $R$, we now need to estimate the extensive part of the free energy. In the states carrying a charged pion condensate, we have $\mathscr{F}_{\text{ext}} \sim \mu_{\text{I}}^2 f_\pi^2$ and accordingly $n_{\text{I}} \sim \mu_{\text{I}} f_\pi^2$ [1]. This estimate is accurate for the uniform BEC state and reasonable also for the Abrikosov vortex lattice except very close to the upper critical field $H_{\text{c2}}$, at which the charged pion condensate vanishes. Combining it with Eq. (A.3) then gives

$$R \lesssim \frac{1}{e f_\pi}. \tag{A.4}$$

To get a similarly simple estimate for the CSL state, we employ the solution valid near the chiral limit, see Sect. 4.1. From Eq. (24) we get at once $\mathscr{F}_{\text{ext}} \sim \Omega^2 H^2$, where the dimensionless parameter $\Omega$ is defined by Eq. (22). Accordingly, $n_{\text{I}} \sim \mathscr{F}_{\text{ext}}/\mu_{\text{I}} \sim \Omega^2 H^2/\mu_{\text{I}}$. This leads to the bound

$$R \lesssim \frac{\mu_{\text{I}}}{e \Omega H}. \tag{A.5}$$

For magnetic fields not too far above the critical field (27), we may in addition write the latter parametrically as $H_{\text{CSL}} \sim m_\pi f_\pi / \Omega$ and use this to simplify the result to

$$R \lesssim \frac{\mu_{\text{I}}}{e m_\pi f_\pi}. \tag{A.6}$$

In the interesting part of the phase diagram where the various ordered phases compete with each other, $\mu_{\text{I}}$ is larger than $m_\pi$. Hence the bound resulting from the free energy of the CSL state is weaker than that from the BEC state, given by Eq. (A.4).

Altogether, we may therefore conclude that the analysis presented in this paper is quantitatively reliable if the system size $R$ falls into the range

$$\frac{1}{m_\pi} \lesssim R \lesssim \frac{1}{e f_\pi}. \tag{A.7}$$

The lower bound herein is needed to ensure that our effective field theory description of the QCD vacuum does not suffer from large finite-volume corrections. Interestingly, this hierarchy of bounds on $R$ suggests that the vacuum pion mass should not be too small, $m_\pi \gtrsim e f_\pi$. The value $m_\pi = e f_\pi$, used in Sect. 6 for numerical illustration, is therefore just on the edge of the domain of validity of our analysis. Pion masses smaller than $e f_\pi$ would require a more careful treatment, explicitly taking into account finite-volume corrections.

## Appendix B: Type-I versus type-II superconductivity

In type-I superconductors, the uniform superconducting condensate is completely destroyed by an external magnetic field given by Eq. (13), above which the normal (vacuum) state prevails. On the other hand, in type-II superconductors, an intermediate phase featuring the Abrikosov lattice of magnetic vortices appears in the range of magnetic fields $H_{\text{c1}} < H < H_{\text{c2}}$. Throughout the present paper, we mostly implicitly assumed that iQCD is a type-II superconductor. In this appendix, we provide some supportive evidence for when type-I or type-II behavior of the charged pion condensate may be expected.

The key observation is that at $H_{\text{c2}}$ derived in Sect. 5.1, the vacuum state becomes unstable with respect to formation of a charged pion condensate. But if $H_{\text{c}} < H_{\text{c2}}$, then the stable equilibrium just below $H_{\text{c2}}$ cannot be the uniform BEC state. It follows immediately that if, for a given fixed value of $\mu_{\text{I}}$, $H_{\text{c2}} > H_{\text{c}}$, then the phase diagram must feature a phase carrying some kind of inhomogeneous pion condensate. It is not a priori given that this condensate is an Abrikosov lattice of vortices. To show that a state carrying vortices is stable and favored over the uniform BEC state, one should rather impose the condition $H_{\text{c1}} < H_{\text{c}}$ [18].

The advantage of our criterion is that it takes a very simple analytic form. Namely, from Eqs. (13) and (34) we obtain immediately that

$$H_{\text{c2}} - H_{\text{c}} = \frac{\mu_{\text{I}}^2 - m_\pi^2}{e}\left(1 - \frac{e f_\pi}{\mu_{\text{I}}}\right). \tag{B.8}$$

Since any kind of charged pion condensate can only be stable if $\mu_{\text{I}} > m_\pi$, we see immediately that a sufficient condition for the existence of a phase with an inhomogeneous pion condensate in the phase diagram is

$$\mu_{\text{I}} > e f_\pi. \tag{B.9}$$

If in addition $m_\pi > e f_\pi$, then some kind of inhomogeneous equilibrium will appear for any value of $\mu_{\text{I}}$ that supports charged pion condensation. On the other hand, if $m_\pi < e f_\pi$, we expect two qualitatively different regimes, depending on whether $\mu_{\text{I}}$ is smaller or larger than $e f_\pi$. For $m_\pi < \mu_{\text{I}} < e f_\pi$, it is in principle possible that as the external magnetic field is cranked up, the uniform charged pion BEC is destroyed in favor of the vacuum. As we explain in Sect. 6.2, we expect in this case a direct first-order transition from the uniform BEC phase to the neutral pion CSL phase.